\documentclass[showpacs,preprintnumbers,amsmath,amssymb,10pt]{revtex4}

\usepackage{graphicx}
\usepackage{dcolumn}
\usepackage{bm}
\usepackage{hyperref}
\usepackage{color}

\definecolor{Blue}{rgb}{0.3,0.3,0.9}
\definecolor{Red}{rgb}{1,0,0}
\definecolor{Green}{rgb}{0,1,0}

\newcommand{\lb}[1]{\label{#1}}
\newcommand{\ff}[1]{(\ref{#1})}

\newcommand{\bark}{\bar{k}}
\newcommand{\barp}{\bar{p}}
\newcommand{\barmu}{\bar{\mu}}
\newcommand{\barn}{\bar{N}}
\newcommand{\epsi}{\epsilon}

\begin{document}
\title{Fully Loop-Quantum-Cosmology-corrected propagation of gravitational waves during
slow-roll inflation}

\author{J. Grain}
\email{julien.grain@ias.u-psud.fr}
\affiliation{%
Institut d'Astrophysique Spatiale, Universit\'e Paris-Sud 11, CNRS \\ B\^atiments 120-121, 91405 Orsay Cedex, France }%

\author{T. Cailleteau}
\email{cailleteau@lpsc.in2p3.fr}
\affiliation{%
Laboratoire de Physique Subatomique et de Cosmologie, UJF, INPG, CNRS, IN2P3\\
53, avenue des Martyrs, 38026 Grenoble cedex, France
}%

\author{A. Barrau}
\email{aurelien.barrau@cern.ch}
\affiliation{%
Laboratoire de Physique Subatomique et de Cosmologie, UJF, INPG, CNRS, IN2P3\\
53, avenue des Martyrs, 38026 Grenoble cedex, France
}%

\author{A. Gorecki}
\email{gorecki@lpsc.in2p3.fr}
\affiliation{%
Laboratoire de Physique Subatomique et de Cosmologie, UJF, INPG, CNRS, IN2P3\\
53, avenue des Martyrs, 38026 Grenoble cedex, France
}%

\date{\today}

\begin{abstract}

The cosmological primordial power spectrum is known to be one of the most promising
observable to probe quantum gravity effects. In this article, we investigate how
the tensor power spectrum is modified by loop quantum gravity corrections. The
two most important quantum terms, holonomy and inverse-volume, are
explicitly taken into account in a unified framework. The equation of propagation of gravitational waves
is derived and solved for one set of parameters.

\end{abstract}


\maketitle

\section{Introduction}

The inflationary scenario is currently the
favored paradigm to describe the first stages of the evolution of the Universe
(see, {\it e.g.}, \cite{linde} for a recent review). Although still debated, 
it has received many experimental confirmations, including from the
WMAP 5-year results \cite{wmap}, and solves most cosmological paradoxes. 

On the other hand, a fully quantum theory of gravity is necessary to 
investigate situations
where general relativity (GR) breaks down. The big bang is an
example of such a situation where the backward evolution of a classical
space-time comes to an end after a finite amount of time. Among the
theories willing to reconcile the Einstein gravity with quantum mechanics,
loop quantum gravity (LQG) is appealing as it is based on a  
nonperturbative quantization of 3-space geometry (see, {\it e.g.}, \cite{rovelli1} 
for an introduction). Loop quantum cosmology (LQC) is a 
finite, symmetry reduced model of LQG  suitable for the study of the
whole Universe as a physical system (see, {\it e.g.}, \cite{bojo0}).

In this article, we consider the influence of LQC corrections to general 
relativity on the production and propagation of  gravitational waves during inflation. 
We first derive the equation of propagation of gravity waves with both holonomy and
inverse-volume corrections. This equation is then reexpressed 
with the {\it commonly used} cosmological variables.
It is finally solved for a specific set of parameters and the primordial power spectrum is derived.
The aim of this work is to conclude our previous studies \cite{grainlqg2} 
and \cite{grainlqg3} where, respectively, only holonomy and only
inverse-volume corrections were considered. By combining both terms, we show that the 
inverse-volume correction dominates over the holonomy one and dictates the overall shape 
of the tensor spectrum.

Quite a lot of work has already been devoted to gravitational waves in LQC \cite{lqcgen}.
Our approach assumes the background to be described by the standard slow-roll
inflationary scenario whereas LQC corrections are taken into account to compute
the propagation of tensor modes. This approach is heuristically justified (to
decouple the physical effects) and intrinsically plausible (as, on the one hand, the LQC-driven
superinflation can only be used to set the proper initial conditions to a
standard inflationary stage if the horizon {\it and} flatness problems are both
to be solved \cite{tsuji} and as, on the other hand, it seems that the quantum bounce can trigger
on a standard inflationary phase \cite{jakub}). In addition, very few studies so far have taken into account
both the holonomy and the inverse-volume corrections. This latter term is somehow more 
speculative than the former one as it was shown to exhibit a fiducial cell dependence 
(see, {\it e.g.}, \cite{ashtekar4}). For the sake of completeness it is however obviously worth
considering the fully corrected propagation of gravitational waves.

\section{Equation of propagation for the graviton}

The derivation of the equation of propagation of gravitational waves with both holonomy
and inverse-volume corrections extensively uses the material developed in \cite{bojo1}: notations, 
conventions and framework of this work are the same and will not be explicitly restated. 
We begin by considering a Friedmann-Lema\^\i{t}re-Robertson-Walker universe with a spatial metric $q_{ab}$ which will be perturbed to
account for
gravitational waves. Hereafter, $N$ and $N^a$ are respectively the lapse function and the 
shift function. The metric components read as follows:

\begin{eqnarray}
g_{00} &=& - N^2 + q_{ab} N^a N^b = -a^2(\eta), \\
g_{0a} &=& q_{ab} N^b = 0,\\
g_{ab} &=& q_{ab} = a^2(\eta) (\delta_{ab} + h_{ab}). \label{metricspatialcor}
\end{eqnarray} 

As usual in the formalism of LQC, we use the Ashtekar variables for an homogeneous and isotropic background: the connection
$\bar{A}_{a}^{i}$, and the triad density $\bar{E}_{i}^{a}$. They can be written as a function of two other variables 
$(\bark, \barp)$ as 
\begin{eqnarray} \label{encodage1}
\bar{E}_{i}^{a} &=& \barp \delta_{i}^{a,}  \nonumber \\ 
\bar{A}_{a}^{i} &=& \bar{K}_{a}^{i} + \bar{\Gamma}_{a}^{i} \nonumber \\
\bar{K}_{a}^{i} &=& \bark \delta_{a}^{i},  \hskip 2truecm \bar{\Gamma}_{a}^{i} = 0, \nonumber \\
\bar{N}^a &=& 0, \hskip 2.4truecm \bar{N} = \sqrt{\barp}.  \label{nna}
\end{eqnarray} 
Hamilton-Jacobi equations will be used to determine the perturbed part of the Ashtekar variables. The Hamiltonian 
constraint reads as
\begin{eqnarray}
H[N] &=&\frac{1}{2\kappa} \int_{\Sigma} d^3 x N |det E|^{-\frac{1}{2}} E_j^a E_k^b (\epsilon_{ijk} F_{ab}^i - 2(1+\gamma^2) K_{[b}^i K_{a]}^j ), \label{Hgr}
\end{eqnarray}
where $F_{ab}^i = \partial_a A_b^i - \partial_b A_a^i + \epsilon^{ijk} A_a^j A_b^k$ is the field strength. The Hamiltonian
for a matter field $\Phi$ is given by
\begin{equation} \label{Hm}
H_{matter} = \int d^3 x \left( \frac{1}{2} \frac{p_\Phi^2 + E_i^a E_i^b \partial_a \Phi \partial_b \Phi}{\sqrt{|det E_j^c|}} +
\sqrt{|det E_j^c|} V(\Phi) \right).
\end{equation}
With Eq.(\ref{encodage1}) and these Hamiltonians, the background is described by 
\begin{equation} \label{Hgravfonda}
H_G^{fond} [\bar{N}] =  \frac{1}{2 \kappa} \int_{\Sigma} d^3 x \bar{N} \left[ -6 \sqrt{\barp} k^2\right] ,
\end{equation} 
and
\begin{equation} \label{Hmatter1}
H_{matter} [\bar{N}] = \int_{\Sigma} d^3 x  \left( \frac{1}{2} \frac{p_{\Phi}^2}{\barp^{\frac{3}{2}}} +
\barp^{\frac{3}{2}} V(\Phi) \right).
\end{equation}
Perturbing the canonical variables (and going through the appropriate Poisson bracket) leads to:
\begin{eqnarray}
H_{\rm G} [\barn] = \frac{1}{2 \kappa}\int_{\Sigma}\mathrm{d}^3x \bar{N} \left[ -6\sqrt{\bar{p}} \bark^2  - \frac{\bark^2}{2\bar{p}^{3/2}} (\delta E^c_j\delta E^d_k\delta_c^k\delta_d^j) + \sqrt{\bar{p}} (\delta K_c^j\delta K_d^k\delta^c_k\delta^d_j) \right. \nonumber \\
\left. - \frac{2 \bark}{\sqrt{\bar{p}}} (\delta E^c_j\delta K_c^j) 
- \frac{1}{\bar{p}^{3/2}} (\delta_{cd} \delta^{jk}  E^c_j \delta^{ef} \partial_e \partial_f  E^d_k ) 
\right],
\end{eqnarray}
where only the tensor perturbations ({\it i.e.} gravitational waves) are considered in $\delta E_i^a$.

This classical Hamiltonian is to be modified by quantum corrections. Because loop quantization is based on
holonomies, {\it i.e.} exponentials of the connection rather than direct connection components, 
one needs to substitute in the gravitational sector
\begin{equation}
\bark \rightarrow \frac{\sin( m \bar{\mu}\gamma\bar{k})}{ m \bar{\mu}\gamma}
\end{equation} 
where $\bar{\mu}$ is a new parameter related to the action of the fundamental Hamiltonian on a lattice
state. In addition, because of inverse powers of the densitized triad which, when quantized, becomes an
operator with zero in the discrete part of its spectrum, the matter and gravitational Hamiltonians must be modified  
by introducing the function
\begin{equation}
\alpha(\barp,\delta E^a_i) = 1 + \lambda q^{n} = 1 + \lambda \left( \frac{l_{PL}^2}{\barp}\right)^{n}.
\end{equation}
At a semiclassical level, {\it i.e.} $q\ll1$, the same parametric form of $\alpha$ can be used in both the matter Hamiltonian and the gravitational
Hamiltonian. However, the two positive and real valued constants $\lambda$ and $n$ may differ from one sector to another. In the following,
$(S,~\lambda_s,~s)$ and $(D,~\lambda_d,~d)$ will therefore denote $(\alpha,~\lambda,~n)$ for the gravitational sector and the matter sector
respectively. With these two corrections, the Hamiltonians read
\begin{eqnarray} \label{egeg}
H_{\rm G}^{\rm eff}[\barn] &=& \frac{1}{2 \kappa}\int_{\Sigma}\mathrm{d}^3x \bar{N} S(\bar{p})
\left[-6\sqrt{\bar{p}} \left(\frac{\sin\bar{\mu}\gamma\bar{k}}{\bar{\mu}\gamma}\right)^2 \right], \\
H_{matter} [\bar{N}] &=& \int_{\Sigma} d^3 x \left( \frac{1}{2} D(q)
\frac{p_{\Phi}^2}{\barp^{\frac{3}{2}}} + \barp^{\frac{3}{2}} V(\Phi) \right), \lb{agag}
\end{eqnarray}
with $H_G^{eff}$ the effective gravitational Hamiltonian describing the homogeneous
background. The equations of motion for $(\bark,\barp)$, {\it i.e.} the background equations, can be obtained in the
Hamiltonian formalism
\begin{equation}
\dot{\barp} = \{ \barp, H_G^{eff} [\bar{N}] + H_{matter} [\barn] \}~;~
\dot{\bark} =  \{ \bark, H_G^{eff} [\bar{N}] + H_{matter} [\barn] \},
\end{equation} 
leading to:
\begin{eqnarray}
\dot{\barp} &=& 2 \cdot \barp \cdot S (\barp ,\delta E) \cdot \left( \frac{sin(2 \barmu \gamma \bark)}{2 \barmu \gamma}\right), 
\lb{dotbarp}  \\
\dot{\bark} &=& \frac{\kappa }{3 V_0} \frac{\partial H_{matter}}{\partial \barp} - \barp \frac{\partial S}{\partial \barp} \cdot
\left( \frac{sin( \bar{\mu} \gamma \bark)}{ \barmu \gamma} \right)^2 - \frac{S}{2} \left[ \left(\frac{sin( \bar{\mu} \gamma \bark)}{
\barmu \gamma} \right)^2  + 2 \barp \frac{\partial}{\partial
\barp} \left(\frac{sin( \bar{\mu} \gamma \bark)}{ \barmu \gamma} \right) ^2 \right]. \lb{dotbark} 
\end{eqnarray}

The same modification is applied to the perturbed gravitational Hamiltonian. Denoting $H_G^{Phen}$ the effective perturbed 
quantum-corrected gravitational Hamiltonian, it reads with both holonomy and inverse-volume corrections
\begin{eqnarray} \label{Hgravpertcor}
H_{\rm G}^{\rm Phen}[N] = \frac{1}{2 \kappa}\int_{\Sigma}\mathrm{d}^3x \bar{N} S(\bar{p},\delta E^a_i)
\left[-6\sqrt{\bar{p}}
\left(\frac{\sin\bar{\mu}\gamma\bar{k}}{\bar{\mu}\gamma}\right)^2
- \frac{1}{2\bar{p}^{3/2}} 
\left(\frac{\sin\bar{\mu}\gamma\bar{k}}{\bar{\mu}\gamma}\right)^2
(\delta E^c_j\delta E^d_k\delta_c^k\delta_d^j) \right. \nonumber\\
+ \left.\sqrt{\bar{p}} (\delta K_c^j\delta K_d^k\delta^c_k\delta^d_j) 
- \frac{2}{\sqrt{\bar{p}}} 
\left(\frac{\sin 2\bar{\mu}\gamma\bar{k}}{2\bar{\mu}\gamma}\right)
(\delta E^c_j\delta K_c^j) 
- \frac{1}{\bar{p}^{3/2}} (\delta_{cd} \delta^{jk}  E^c_j \delta^{ef} 
\partial_e \partial_f  E^d_k ) 
\right].
\end{eqnarray}
We now turn to the equation of motion of the graviton. The perturbed densitized triad is
\begin{equation} \lb{pertdenstriad}
\delta E_i^a = - \frac{1}{2} \barp h_i^a.
\end{equation}
As has been done for the homogeneous canonical variables, it is possible to define the equation of motion for
the perturbations:
\begin{eqnarray}
\delta \dot{E_i^a} &=& \{ \delta E_i^a , H_G^{Phen} [\bar{N}] + H_{matter} [\bar{N}] \} \nonumber \\
                &=&  - \{\delta K_b^j (x),\delta E_i^a(y)\} \frac{\delta}{{\delta (\delta K_b^j)}} (H_G^{Phen} [\bar{N}] + H_{matter}
[\bar{N}]), \nonumber \\
\delta \dot{K_a^i} &=& \{ \delta K_a^i , H_G^{Phen} [\bar{N}] + H_{matter} [\bar{N}] \}  \nonumber\\
                &=&    \{\delta K_a^i(x),\delta E_j^b(y)\} \frac{\delta}{\delta (\delta E_j^b)} ( H_G^{Phen} [\bar{N}] + H_{matter}
 [\bar{N}]) . \nonumber
\end{eqnarray}
This leads to:
\begin{eqnarray} 
\delta \dot{E}_i^a &=& - \frac{1}{2} ( \dot{\barp} h_i^a + \barp \dot{h}_i^a) \lb{derivE}\\
&=& - S (\barp, \delta E) \cdot \left[\barp \cdot\delta K_c^l \cdot \delta_a^c \cdot  \delta_i^b - 
\left(\frac{sin(2 \barmu \gamma \bark)}{2 \barmu \gamma} \right) \cdot \delta E_i^a \right]. \lb{deltaK}
\end{eqnarray}
By combining those equations and using the expression of $\dot{\barp}$, one obtains the expression of 
$\delta K_a^i$ as a function of $h_a^i$ and of $\dot{h}_a^i$. The expression of $\delta K_a^i$
is:
\begin{equation}
\delta K_a^i = \frac{1}{2 S} \dot{h}_a^i + \frac{1}{2} \left( \frac{sin(2 \barmu \gamma \bark)}{2 \barmu \gamma} \right)
h_a^i.
\end{equation}
The equation of motion  will lead to another derivative with respect to $\eta$. 
The Hamilton-Jacobi equation for the perturbed connection can now be used to find the final equation of propagation for gravitational 
waves:
\begin{eqnarray}
\delta \dot{K_a^i} &=& \frac{1}{2} \left[ \frac{\ddot{h}_a^i}{S}  - \frac{1}{S^2} \frac{\partial S}{\partial \eta} \cdot
\dot{h}_a^i + \left( \frac{sin(2 \barmu \gamma \bark)}{2 \barmu \gamma}\right) \dot{h}_a^i + h_a^i \cdot \frac{\partial}{\partial \eta} 
\left( \frac{sin(2 \barmu \gamma \bark)}{2 \barmu \gamma}\right)  \right]  \nonumber\\
                &=&\{\delta K_a^i(x),\delta E_j^b(y)\} \frac{\delta}{\delta (\delta E_j^b)} ( H_G^{Phen} [\bar{N}] + H_{matter}
 [\bar{N}]). \nonumber
\end{eqnarray}
As 
\begin{eqnarray*}
\frac{\delta H_G^{Phen}}{\delta (\delta E_j^b)} &=& \frac{1}{2 \kappa} \int_{\Sigma} d^3 (x) \cdot \barn \cdot  
\frac{\delta S}{\delta (\delta E_j^b)} [...] + \frac{1}{2 \kappa} \int_{\Sigma} d^3 (x) \barn S \left[ -\frac{2}{2 \barp^{\frac{3}{2}}}
\left(\frac{sin(\barmu \gamma \bark)}{\barmu \gamma} \right)^2
(\delta E_l^c \cdot \delta_b^l \cdot \delta_c^j) \right.\\
&-& \left. \frac{2}{\sqrt{\barp}} \left( \frac{sin(2 \barmu \gamma \bark)}{2 \barmu \gamma}
\right) \delta K_b^j - \frac{2}{\barp^{\frac{3}{2}}} (\delta_{bd} \cdot \delta^{jk} \cdot \delta_{ef} \partial_e \partial_f (\delta
E_k^d))\right],
\end{eqnarray*}
where [...] stands for the term beginning with$ [-6\sqrt{\bar{p}}
\left(\frac{\sin\bar{\mu}\gamma\bar{k}}{\bar{\mu}\gamma}\right)^2 - ...]$ in \ff{Hgravpertcor}, one
obtains (with $\delta^{ef} \partial_e \partial_f (\delta E_k^d) =  \nabla^2 (\delta E_k^d) =-
\frac{1}{2} \barp \cdot\nabla^2 h_k^d$)
\begin{eqnarray}
\{\delta K_a^i,\delta E_j^b\}\frac{\delta H_G^{Phen}}{\delta (\delta E_j^b)} 
&=& \frac{1}{2} \sqrt{\barp} 
\frac{\delta S}{\delta (\delta E_j^b)} [...] \nonumber\\
&+& \frac{1}{2} S \left[ \frac{1}{2} \left(\frac{sin(\barmu \gamma \bark)}{\barmu \gamma} \right)^2 h_a^i - 
\left( \frac{sin(2 \barmu \gamma \bark)}{2 \barmu \gamma}\right) \left(\frac{\dot{h}_a^i}{S} + \left(\frac{sin(2 \barmu \gamma
\bark)}{2 \barmu \gamma}\right) h_a^i\right) + \nabla^2 h_a^i \right] \lb{63} \\
&=& \frac{1}{2} \left[ \frac{\ddot{h}_a^i}{S}  - \frac{1}{S^2} \frac{\partial S}{\partial \eta}
\dot{h}_a^i + \left( \frac{sin(2 \barmu \gamma \bark)}{2 \barmu \gamma}\right) \dot{h}_a^i + h_a^i \frac{\partial}{\partial \eta} 
\left( \frac{sin(2 \barmu \gamma \bark)}{2 \barmu \gamma}\right)  \right]  - \kappa \frac{\delta H_{matter} [\barn]}{\delta (\delta E_j^b)}.
\lb{64}
\end{eqnarray}

After quite a lot of algebra, the equation of motion of the graviton can be derived:
\begin{equation} \label{equapropagation}
\frac{1}{2} \left[ \ddot{h}_a^i + 2 S \left( \frac{sin(2 \bar{\mu} \gamma \bark)}{2 \barmu \gamma} \right) \dot{h}_a^i
\left(1-\frac{\barp}{S} \frac{\partial S}{\partial \barp} \right) - S^2 \nabla^2 h_a^i + S^2 T_Q h_a^i \right] +
S \mathcal{A}_a^i = \kappa S \Pi_{Q_a}^i,
\end{equation}
where
\[
T_Q = - 2 \left( \frac{\barp}{\barmu} \frac{\partial \barmu}{\partial \barp} \right) (\barmu \gamma)^2 \left( \frac{sin( \bar{\mu} \gamma
\bark)}{\barmu \gamma} \right)^4,
\]

\[
\Pi_{Q_a}^i = \frac{1}{3 V_0} \frac{\partial H_{matter}}{\partial \barp} \left( \frac{\delta E_j^c \delta_a^j
\delta_c^i}{\barp} \right) cos (2 \barmu \gamma \bark) + \frac{\delta H_{matter}}{\delta ( \delta E_i^a)},
\]

\[
\mathcal{A}_a^i = \frac{1}{2} \sqrt{\barp} \frac{\delta S}{\delta(\delta E_i^a)} [...] - \barp \frac{\partial S}{\partial
\barp} cos(2 \barmu \gamma \bark) \left( \frac{sin( \bar{\mu} \gamma \bark)}{\barmu \gamma} \right)^2
h_a^i.
\]
As usual, requiring an anomaly-free constraint algebra in the presence of quantum corrections requires
$\mathcal{A}_a^i $ to vanish. It should be noticed that the inverse-volume correction is
involved in each term, through the $S$ and $D$ factors, whereas the holonomy correction is only involved in the 
$\dot{h}_a^i$ term, in  $T_Q$ and in $\Pi_{Q_a}^i$.

It is worth studying a bit more into the details of this
$\Pi_{Q_a}^i$ source term as it seems to have been misunderstood in several works. In particular, it has
often been either neglected or miscomputed. Without holonomy and inverse-volume correction, this term reads as
\begin{equation}
\Pi^i_a = \left[ \frac{1}{3 V_0} \frac{\partial H_{matter}} {\partial \barp} \left( \frac{\delta E_j^c \delta_a^j \delta_c^i}{ \barp} \right)
+ \frac{\delta H_{matter}}{\delta ( \delta E_i^a)} \right],
\label{class-source}
\end{equation}
with, in this case,
\begin{equation}
E_i^a = \barp \delta_i^a ~,~
\delta E_i^a = - \frac{1}{2} \barp h_i^a~,~ 
det{E} = \frac{1}{3 !} \epsi_{abc} \epsi^{ijk} E_i^a E_j^b E_k^c.
\end{equation}
At the zeroth order in gravitational perturbation, one can show that
\begin{equation}
\bar{H}_{matter} = \int_{\Sigma} d^3 x \barn \left( \frac{1}{2}\frac{p_\phi^2}{\bar{p}^{\frac{3}{2}}} + \bar{p}^{\frac{3}{2}}V(\phi) \right),
\end{equation}
and the nonlinear $H_{matter}$ is given by
\begin{equation}
H_{matter} = \bar{H}_{matter} + \int_{\Sigma} d^3 x \barn \frac{1}{4 \sqrt{\barp}} \delta E_i^a \delta E_j^b \delta_a^j \delta_b^i \left(
\frac{1}{2} \frac{p_{\phi}^2}{\barp^3} -  V(\phi) \right),
\end{equation}
thus leading to
\begin{equation}
\frac{\delta H_{matter}}{\delta (\delta E_i^a)} = \frac{\barn}{2} \frac{\delta E_j^b}{\sqrt{\barp}} \delta_a^j \delta_b^i \left( \frac{1}{2}
\frac{p_{\phi}^2}{\barp^3} - V(\phi) \right).
\end{equation}
Restricting to the first order in perturbation, the derivative with respect to $\barp$ can be evaluated and one finally obtains
\begin{equation}
\frac{1}{3 V_0} \frac{\partial H_{matter}} {\partial \barp}  \frac{\delta E_j^c \delta_a^j \delta_c^i}{ \barp} = -
\frac{\delta H_{matter}}{\delta ( \delta E_i^a)}.
\end{equation}
This easily establishes that classically $\Pi^i_a=0$. However, when LQC corrections are taken into account
the source term may not vanish anymore (because of the derivative of $D$ with respect to $\bar{p}$ for the inverse-volume correction and because of the cosine term for the holonomy one). 

When only inverse-volume corrections are considered, the source term is still given by Eq. (\ref{class-source}) but the matter Hamiltonian now reads 
\begin{eqnarray}
H_{matter} &=&\bar{H}_{matter}+H^{(\delta)}_{matter} \\
&=& \int_{\Sigma} d^3 x \barn\left[\left(D(\bar{p},\delta E^a_i)\frac{1}{2}\frac{p_\phi^2}{\bar{p}^{\frac{3}{2}}} + \bar{p}^{\frac{3}{2}}V(\phi)
\right)+\frac{1}{4 \sqrt{\barp}} \delta E_i^a \delta E_j^b \delta_a^j \delta_b^i \left(D(\bar{p},\delta E^a_i)
\frac{1}{2} \frac{p_{\phi}^2}{\barp^3} -  V(\phi) \right)\right], \nonumber
\end{eqnarray}
which leads, at the leading order, to
\begin{equation}
\frac{\delta H_{matter}}{\delta (\delta E_i^a)} = \barn \left[ \frac{\delta E_j^b}{2\sqrt{\barp}} \delta_a^j \delta_b^i \left( \frac{1}{2}
\frac{p_{\phi}^2}{\barp^3} - V(\phi) \right) + \frac{p_{\phi}^2}{2 \barp^{\frac{3}{2}}}  \frac{\delta D}{\delta (\delta E_i^a)} \right],
\end{equation}
and
\begin{equation}
\frac{1}{3 V_0} \frac{\partial H_{matter}} {\partial \barp}  \frac{\delta E_j^c \delta_a^j \delta_c^i}{ \barp} = \barn \frac{1}{3}
\left(\frac{\delta E_j^c \delta_a^j \delta_c^i}{\barp}\right) \left[ -\frac{3}{4} \frac{D}{\barp^{\frac{5}{2}}} p_{\phi}^2 + \frac{3}{2}
\sqrt{\barp} V(\phi) + \frac{\partial D}{\partial \barp} \frac{p_{\phi}^2}{2} \frac{1}{\barp^\frac{3}{2}} \right].
\end{equation}
We finally obtain
\begin{equation}
\Pi_{Q_a}^{i,(IV)} = \frac{1}{3 V_0} \left(\frac{\delta E_j^c \delta_a^j \delta_c^i}{ \barp} \right) \frac{\partial H_{matter}} {\partial \barp}  +
\frac{\delta H_{matter}}{\delta ( \delta E_i^a)} = \frac{p_{\phi}^2}{2 \barp^\frac{3}{2}} \left[
\frac{1}{3} \left(\frac{\delta E_j^c \delta_a^j \delta_c^i}{\barp} \right) \frac{\partial D}{\partial \barp  } + 
\frac{\delta D}{\delta(\delta E_i^a)}   \right].
\end{equation}
However, because of the anomaly-free condition (see Eq. (27) of \cite{bojo1}), this term is vanishing. This means that, at the leading order, $\Pi_{Q_a}^{i,(IV)}=0$.

Considering now the holonomy correction alone, one can expand the cosine term in $\Pi_{Q_a}^i$ and show that
\begin{eqnarray}
\Pi_{Q_a}^{i,(holo)}= -2 \barmu\gamma\sin^2\left(\barmu\gamma\bark\right)\frac{1}{3 V_0} \frac{\partial \bar{H}_{matter}}{\partial \barp} \left( \frac{\delta E_j^c \delta_a^j
\delta_c^i}{\barp} \right).
\nonumber
\end{eqnarray}

Considering simultaneously the two types of corrections and using the explicit expression of the matter Hamiltonian, one obtains the full LQC source term:
\begin{eqnarray}
\Pi_{Q_a}^{i,(LQC)}=-2 \barmu\gamma\sin^2\left(\barmu\gamma\bark\right)\frac{1}{3 V_0} \frac{\partial \bar{H}_{matter}}{\partial \barp} \left( \frac{\delta E_j^c \delta_a^j
\delta_c^i}{\barp} \right),
\end{eqnarray}
as expected from the vanishing inverse-volume source term.

\section{Schr\"odinger equation for the Fourier modes}

The energy density and the pressure of the cosmological fluid can be written as
\begin{equation} \lb{energydensitydef}
\rho =  \frac{1}{V_0 \barp^{\frac{3}{2}}} \frac{\delta  H_{matter}}{\delta \bar{N}}~,~
p = - \frac{1}{\bar{N} V_0} \frac{\delta H_{matter}[\bar{N}]}{\delta (\sqrt{|det{E}|})}. 
\end{equation}
With the Hamiltonian constraint, one obtains
\begin{equation}
0 = \frac{1}{2 \kappa} \int_{\Sigma} d^3 x S \left[-6 \sqrt{\barp} \left(\frac{sin(\barmu \gamma \bark)}{\barmu \gamma} \right)^2 
\right]
+ \frac{\delta  H_{matter}}{\delta \bar{N}},
\end{equation}  
which finally leads to
\begin{equation}
\rho 
= \frac{3}{\kappa} \frac{S}{\barp}  \left(\frac{sin(\barmu \gamma \bark)}{\barmu \gamma} \right)^2. \lb{energydensity1}
\end{equation}
Defining $\mathcal{H}$ as the Hubble parameter with respect to the conformal time
$(\mathcal{H}=a^{-1}da(\eta)/d\eta))$, we obtain the quantum Friedmann equations:
\begin{eqnarray}
\mathcal{H}^2 & = & S^2 \left(\frac{sin( 2 \barmu \gamma \bark)}{ 2 \barmu \gamma} \right)^2 \nonumber \\
& =& S^2  \frac{\barp}{S} \frac{\kappa}{3} \rho \left(1- \barmu^2 \gamma^2 \frac{\kappa}{3} \frac{\barp}{S} \rho \right) ,
\lb{Fried1}
\end{eqnarray}
which lead, with $\rho_c = 3/(\kappa \barmu^2 \gamma^2 \barp)$, to
\begin{equation} \lb{Friedman}
\mathcal{H}^2 = a^2 \frac{\kappa}{3} \rho \left(S- \frac{\rho}{\rho_c} \right).
\end{equation}
This equation, which has already been found in \cite{calcagni}, includes all the LQC corrections and
shows that the holonomy term, leading to the bounce, is the most important one as far as the background in
concerned. This conclusion will be radically modified for perturbations. 

The equation of motion
for the graviton can now be reexpressed in terms of the commonly-used cosmological variables. By taking into account 
Eq.~\ff{energydensity1} and $\barmu^2 \barp = l_{PL}^2$, one obtains
\begin{equation}
S^2 T_Q 
= - 2 \left( \frac{\barp}{\barmu} \frac{\partial \barmu}{\partial \barp} \right) (S \barmu \gamma)^2 
\left( \frac{sin( \bar{\mu} \gamma \bark)}{\barmu \gamma} \right)^4=\frac{\kappa}{3} \frac{a^2}{\rho_c} \rho^2.
\end{equation}
The multiplicative factor of $\dot{h}_a^i$ in Eq. (\ref{equapropagation}) can be reexpressed as a function of the Hubble parameter
\begin{equation}
2 S  \left( \frac{sin(2 \bar{\mu} \gamma \bark)}{2 \barmu \gamma} \right) \left(1-\frac{\barp}{S} 
\frac{\partial S}{\partial \barp} \right) = 2 \mathcal{H} \left(1-\frac{1}{2} 
\frac{a}{\dot{a}} \frac{\dot{S}}{S} \right).
\end{equation} 
Finally, the source term can be explicitly computed
\begin{equation}
\Pi_{Q_a}^i = \frac{h_a^i}{ S}\frac{\rho}{\rho_c} \frac{\barp}{2} \left[ \rho -\frac{\dot{\Phi}^2}{D(q) a^2} \left( 1- \frac{1}{6}
\frac{\dot{D}}{D} \frac{a}{\dot{a}} \right)\right]. \nonumber
\end{equation}
As in \cite{bojo1}, we use the effective parametrization $S=1 + \lambda_s (q)^{-\frac{s}{2}}$ with
$q=(a/l_{PL})^2$. The
equation of propagation can now be written as
\begin{equation} \lb{propag1}
\ddot{h} +  2 \frac{\dot{a}}{a} \left(1-\frac{1}{2} \frac{\partial \ln(S)}{\partial \ln(a)} \right)  \dot{h} - 
\left(S^2 \nabla^2 + M^2(a) \right)h =0,
\end{equation}
with
\begin{equation}
M^2(a) = \kappa 
\frac{\rho}{\rho_c}  a^2 \left( \frac{2}{3}\rho -\frac{\dot{\Phi}^2}{D(q) a^2} \left( 1- \frac{1}{6} \frac{\dot{D}}{D} \frac{a}{\dot{a}}
\right)\right).
\end{equation}
This can be usefully expressed as an equation for the spatial Fourier transform $h_k$ of $h$
\begin{equation} \lb{equafour}
\ddot{h}_k + 2 \frac{\dot{a}}{a} \left(1- \frac{1}{2} \frac{a}{\dot{a}} \frac{\dot{S}}{S} \right) \dot{h}_k + (S^2 k^2 - M^2(a)) h_k = 0.
\end{equation}
The variables are changed according to $\phi_k = h_ka/\sqrt{S}$, leading to a Schr\"{o}dinger-like
equation
\begin{equation} \lb{equafinale}
\ddot{\phi}_k+\left\{S^2 k^2- \left(\frac{\ddot{a}}{a}+M^2(a) -\frac{\dot{a}}{a} \frac{\dot{S}}{S}+ \frac{3}{4}\left(\frac{\dot{S}}{S}
\right)^2  -\frac{1}{2}\frac{\ddot{S}}{S}\right) \right\}\phi_k=0.
\end{equation}

\section{Power spectrum}

The main question to address is to investigate if one correction, either holonomy or inverse-volume,
dominates over the other as far as the production of gravitational waves during inflation is concerned.
The system describing the dynamics is
\begin{eqnarray*}    
\mathcal{H}^2 &=& a^2\frac{\kappa}{3} \rho \left(S - \frac{\rho}{\rho_c} \right) ,\\
0 &=&\ddot{\Phi}_k + 2\frac{\dot{a}}{a}\left(1-\frac{1}{2}\frac{a}{\dot{a}} \frac{\dot{D}}{D} \right)\dot{\Phi}_k + a^2 D V_{,\Phi}(\Phi), \\
0&=& \ddot{\phi}_k+\left\{{S}^2 k^2- \left(\frac{\ddot{a}}{a}+M^2(a) -\frac{\dot{a}}{a} \frac{\dot{S}}{S}+
\frac{3}{4}\left(\frac{\dot{S}}{S}
\right)^2  -\frac{1}{2}\frac{\ddot{S}}{S}\right) \right\}\phi_k,
\end{eqnarray*}
which is unfortunately much too difficult to be analytically solved. We therefore turn to the approach
developed in \cite{grainlqg2,grainlqg3}. The background evolution is assumed to be classical ($D\approx
1$) with the
scale factor given by the usual slow-roll approximation $a(\eta) = l_0 |\eta|^{-1-\epsi}$. In this case,
the effective Schr\"{o}dinger equation $\left[ \frac{d^2}{d \eta^2} + E_k(\eta) - V(\eta)   \right] \phi_k(\eta) = 0,$
reads, to first order in $\lambda_s$, as
\begin{eqnarray}
E_k(\eta) &=& S^2 k^2 = \left[1+ 2\lambda_s \left( \frac{l_{PL}}{l_0} \right)^s |\eta|^{s(1+\epsi)} \right] k^2, \\
V(\eta) &=& \frac{2+3 \epsi}{\eta^2} +\frac{6}{\kappa} \frac{1}{\rho_c}  \frac{(1+4 \epsi)}{l_0^2} |\eta|^{-2 (1-\epsi)}  \nonumber \\
&+& \lambda_s \left(\frac{l_{PL}}{l_0} \right)^s  \left[ - \frac{12}{\kappa} \frac{1}{\rho_c} \frac{(1+4 \epsi)}{l_0^2}
|\eta|^{s-2 + \epsi (s+2)} + s(1+2\epsi) |\eta|^{s(1+\epsi)-2} - \frac{1}{2} s(s-1+\epsi(2s-1)) |\eta|^{s(1+\epsi)-2} \right]. \nonumber \\
\end{eqnarray}
To implement initial conditions, we consider the limit $\eta \rightarrow -\infty$ where the adiabatic vacuum holds. 
Of course,
if higher order terms in $\lambda_s$ were to be included, the vacuum would not be the same anymore. However, we have checked that the adiabaticity condition
would still be fulfilled in the relevant wavenumber range.

It is possible to solve analytically this equation, at least for one set of parameters: $s=2$ and
$\epsilon=0$. It becomes
\begin{equation} \lb{eqinit}
\frac{d^2 \phi_k}{d \eta^2} + \left[ \left(1+ 2\lambda_s \left( \frac{l_{PL}}{l_0} \right)^2 \eta^2 \right) k^2  - 
\frac{2}{\eta^2} \left(1-\frac{3}{\kappa} \frac{1}{\rho_c}  \frac{1}{l_0^2} \right) - 
\lambda_s \left(\frac{l_{PL}}{l_0} \right)^2  \left[ - \frac{12}{\kappa} \frac{1}{\rho_c} \frac{1}{l_0^2}
+1 \right] \right] \phi_k = 0.
\end{equation} 
By some appropriate changes of variables, this equation can be turned into a Whittaker equation. The solution
can be expressed with Kummer functions and the Wronskian condition 
$\phi_k \partial_\eta \phi_k^{+} - \phi_k^{+} \partial_\eta \phi_k =  16i  \pi/M_{PL}^2$ allows one to
normalize the modes. The field is then given at the end of inflation by
\begin{equation}
\phi_k (c)  = \frac{2 \sqrt{2 \pi}}{M_{PL} (k \sqrt{2Z})^{\frac{1}{4}}}  e^{\frac{i}{2} \pi a} e^{-\frac{i}{2} c} c^{\frac{1}{4}+ \mu}
U \left( \frac{1}{2} + \mu - v , 1 + 2 \mu, i c \right),
\end{equation}
and the resulting primordial tensor power spectrum is
\begin{equation} \lb{spectrefinal}
P_T (k) = \frac{16}{M_{PL}^2}  k^{3-2\mu} H_0^2 (\sqrt{2Z})^{-2\mu} \left| \frac{\Gamma(b-1)}{\Gamma(a)} e^{-\frac{i}{2} \pi
v}\right|^2,
\end{equation}
with 
\begin{eqnarray}
a &=& \frac{1}{2} + \mu - v = \frac{1}{2} + \frac{3}{4} \sqrt{1 + \frac{8 \gamma^2l^2_{PL}}{9 l_0^2}} + \frac{i}{\sqrt{32 Z k^2}} 
\left(k^2 - Z\left(1 - 4 \frac{\gamma^2l^2_{PL}}{l_0^2} \right) \right) \lb{notthescalefactor},\\ 
b &=& 1 + 2 \mu = 1 + \frac{3}{2} \sqrt{1 + \frac{8 \gamma^2l^2_{PL}}{9 l_0^2}}, \\
v &=& -\frac{i}{\sqrt{32 Z k^2}} \left(k^2 - Z\left(1 - 4 \frac{\gamma^2l^2_{PL}}{l_0^2} \right) \right),
\end{eqnarray}
where $Z=(l_{PL}/l_0)^2\lambda_s$ and $\gamma^2=3/(\kappa\rho_c l_{PL}^2)$.
The ultraviolet limit of this spectrum can be easily derived and leads to
\begin{equation}
P_T^{UV} (k) = 16 \pi^3 \left(\frac{l_{PL}}{l_0^2} \right)^2 \left( 1 + \frac{3}{2} \frac{Z}{k^2} (1 - 4 \epsi)\right) k^{-
\frac{4}{3} \omega},
\end{equation}
with $\omega=\gamma^2l^2_{PL}/l_0^2$. On the other hand, the infrared limit is given by
\begin{equation}
P_T^{IR}(k) = 16 \pi^3 \left(\frac{l_{PL}}{l_0} \right)^2 (Z(1-4\omega))^{-\frac{3}{2}} k^3 e^{\pi \sqrt{\frac{Z}{8}}
\frac{(1-4\omega)}{k}}.
\end{equation}
Those results show that the $ k \rightarrow + \infty$ limit of the power spectrum is in agreement with the
general relativistic behavior with the addition of a slight tilt. The ultraviolet spectrum is nearly asymptotically scale invariant. This is not
surprising as both the holonomy correction (encoded in the $k^{-
\frac{4}{3} \omega}$ term) and the inverse-volume correction (encoded in the $ \left( 1 + \frac{3}{2} \frac{Z}{k^2} (1 - 4 \epsi)\right)$ term), 
taken individually, lead to this
behavior. The infrared limit is more interesting as, in this case, the holonomy and inverse-volume
corrections lead to very different spectra. The result obtained here shows that the power spectrum is
exponentially divergent, in exact agreement with the limit obtained with the inverse-volume correction alone.
This proves that, under the standard inflationary background evolution hypothesis, the inverse-volume
term strongly dominates over the holonomy one. This is to be contrasted with the background evolution in the
very remote past where the holonomy term alone leads to the replacement of the singularity by a bounce.

\section{conclusion}

This work derives the fully LQC-corrected equation of motion for gravitational waves. This equation is
expressed in terms of cosmological variables and is explicitly solved for a given set of parameters in
a standard inflationary background. It is shown that the spectrum remains exponentially infrared
divergent, as for a pure inverse-volume correction. This reinforces the use of primordial 
gravitational waves as a strong probe of loop quantum gravity effects. The next step is naturally to
build a fully consistent model which includes all the corrections for both the perturbations and the
background . 

\acknowledgments

This work was supported by the Hublot - Gen\`eve company.

\end{document}